# The Entropies

Roumen Tsekov

Department of Physical Chemistry, University of Sofia, 1164 Sofia, Bulgaria

Entropy is critically examined as a fundamental concept in contemporary science and informatics. Although the typical Shannon entropy provides a proper framework for describing the canonical ensemble, it fails to represent adequately the microcanonical ensemble. This discrepancy manifests additionally in its inability to support a theoretical derivation of the Second Law of thermodynamics.

Clausius introduced entropy as a fundamental quantity of thermodynamics in 1865, while Boltzmann discovered its probabilistic nature 10 years later. According to the Second Law of thermodynamics, entropy drives the arrow of time in isolated systems. At present the entropic concept appears all over the Science and it is especially important nowadays to the modern quantum computers, artificial intelligence and informatics, because information is simply entropy change (Yaglom, Yaglom, 1983). In information theory mathematical definitions of entropy are proposed and perhaps the most general one is given by Rényi (Rényi 1961)

$$H_\alpha \equiv \ln \sum p_i^\alpha /(1-\alpha) \tag{1}$$

where $p_i$ is the probability for observation of a given random event. As is seen, the Rényi formula (1) describes a class of entropy functions via the free positive parameter $\alpha \geq 0$. For example, the Hartley function $H_0$ gives the maximal entropy, $H_1 = -\sum p_i \ln p_i$ is Shannon's entropy (Shannon 1948), while $H_2 = -\ln \sum p_i^2$ is known as collision entropy. Expanding the logarithmic term from Eq. (1) yields the Tsallis entropy (Tsallis 1988), which is widely used for description of non-additive systems, e.g. the living ones,

$$\widetilde{H}_\alpha = (\sum p_i^\alpha - 1)/(1-\alpha) \tag{2}$$

In particular $\widetilde{H}_1 = H_1$, while the linear entropy $\widetilde{H}_2 = 1 - \sum p_i^2$ is adapted as a measure for purity of quantum states. Pareto's distribution, being popular in econophysics (Rosser 2021), maximizes Tsallis' entropy $\widetilde{H}_\alpha$ at constant energy for the case of $\alpha < 1$. As is seen, there are many entropies and they are different.

Returning to physics, let us consider an ideal gas whose particles do not interact with each other. In the frames of Boltzmann's statistics $\{n_i\}$ are the occupation numbers of gas particles on existing energy levels $\{\epsilon_i\}$. Hence, the total number of particles equals to $N = \sum n_i$, while the system energy $E = \sum n_i \epsilon_i$ is the true one, not an average energy from statistical considerations. Boltzmann introduced the multiplicity $W \equiv N!/\prod n_i!$, which is counting the combinations of the $N$ particles among different energy levels. Note that permutations on a single level are physically indistinguishable because no energy change is involved. Ingeniously he related $W$ to the Boltzmann entropy

$$S_B \equiv k_B \ln W = -N k_B \sum f_i \ln f_i \tag{3}$$

where the Stirling formula is employed in the derivation of the last expression. Interpreting $f_i \equiv n_i/N$ as posterior probability for a particle to occupy a given level, Eq. (3) becomes proportional to the Shannon entropy, and Boltzmann's constant $k_B$ just scales the information entropy to the physical one. Because the full probability $\prod f_i^{n_i} = 1/W$ of independent gas particles equals the reciprocal statistical weight, Eq. (3) can be alternatively interpreted as Shannon's expression for a uniform full probability, i.e. Hartley's function. Furthermore, Boltzmann proved the H theorem stating that $S_B$ continuously grows over time until it reaches a maximum corresponding to the thermodynamic equilibrium. So, he deduced the Second Law. Maximizing $W$ at constant energy $E$, volume $V$ and number of particles $N$ yields the Boltzmann distribution $f_i = \exp(-\beta \epsilon_i)/Q$, where $\beta \equiv 1/k_B T$ is the reciprocal temperature and $Q \equiv \sum \exp(-\beta \epsilon_i)$ provides the normalization $\sum f_i = 1$. The equilibrium Helmholtz free energy $F \equiv E - TS = -N k_B T \ln Q$ follows by substitution of Boltzmann's distribution in Eq. (3), and one can easily check now the validity of the Gibbs-Helmholtz equation $E = \partial(\beta F)/\partial \beta$. The equilibrium entropy can be also calculated via $S = -(\partial F/\partial T)_{V,N}$ and the result is the well-known Sackur-Tetrode equation (Huang 1991).

The Boltzmann theory is restricted to ideal gases of non-interacting particles. To describe real gas, one should consider it as a mechanical system defined by the Hamilton function $H(p,q)$, where $p$ and $q$ are the $3N$-dimensional vectors of particles' momenta and positions, respectively. Naturally, the Hamilton function depends parametrically on the number of particles $N$ and volume $V$, but not on energy and temperature. In the frames of statistical mechanics, the system is described by the non-equilibrium probability density $\rho(p,q,t)$, which is normalized by definition,

$\int \rho d\Gamma = 1$. Entropy is traditionally associated in statistical mechanics with the Gibbs-Shannon expression, because it is additive and clear functional on the probability density $\rho$,

$$S_G \equiv -k_B \int \rho \ln \rho \, d\Gamma \qquad (4)$$

The dimensionless infinitesimal phase space volume $d\Gamma \equiv d^{3N}p \, d^{3N}q / N! \, h^{3N}$ is reduced by the Planck constant $h$ for each action degree, while $N!$ accounts the indistinguishability of the particles. A plausible extension $S_G = -k_B tr(\hat{\rho} \ln \hat{\rho})$ is proposed for quantum systems via the density matrix operator $\hat{\rho}(t)$ (von Neumann 1955).

Let us consider first a real system exchanging heat with surroundings. Thus, energy is not integral of motion, but temperature $T$ stays constant. It is well known that in this case the canonical Gibbs distribution $\rho_{eq} = \exp(-\beta H)/Z$ is the equilibrium probability density (Huang 1991), where the canonical partition function $Z(T,V,N) \equiv \int \exp(-\beta H) d\Gamma$ follows from normalization. For ideal gas the canonical Gibbs distribution naturally decomposes to a product of the individual Boltzmann's distributions. Substituting $\rho_{eq}$ in Eq. (4) yields the equilibrium Gibbs entropy

$$S_G = U/T + k_B \ln Z \qquad (5)$$

where $U \equiv \int H \rho_{eq} d\Gamma$ is the internal energy of the system. Hence, Eq. (5) reveals the equilibrium Helmholtz free energy $F = -k_B T \ln Z$, which is the characteristic function at constant temperature, volume and number of particles. Hence, one can calculate the conjugated thermodynamic parameters such as pressure, chemical potential and entropy, respectively,

$$p \equiv -(\partial F/\partial V)_{T,N} = -\int (\partial H/\partial V)_N \rho_{eq} d\Gamma \qquad (6)$$

$$\mu \equiv (\partial F/\partial N)_{T,V} = \int (\partial H/\partial N)_V \rho_{eq} d\Gamma \qquad (7)$$

$$S \equiv -(\partial F/\partial T)_{V,N} = U/T + k_B \ln Z \qquad (8)$$

It can be shown either in classical or in quantum statistical mechanics (Tsekov 2021) that the non-equilibrium free energy decreases continuously over time until it reaches a minimum at equilibrium as required by the Second Law of thermodynamics. Therefore, everything looks perfect in isothermal systems, and the obvious reason is that the exponent in the canonical Gibbs distribution is just the inverse function of the logarithm in Eq. (4).

Let us consider now an isolated real system, where the energy $E$ is constant. It is known that in this case the equilibrium probability density is given by the microcanonical Gibbs distribution $\rho_{eq} = \delta(E - H)/\Omega$, where $\delta$ is the Dirac delta function and the normalization constant reads $\Omega \equiv \int \delta(E - H) d\Gamma$ (Huang 1991). In quantum mechanics $\Omega$ is proportional to degeneracy of the corresponding energy level $E$ of the system. Substitution of $\rho_{eq}$ in Eq. (4) yields infinitely negative entropy $S_G = k_B \ln[\Omega/\delta(0)]$, which is obviously unphysical. To derive the correct expression for entropy of isolated systems one should calculate again the pressure and chemical potential

$$p \equiv -\int (\partial H/\partial V)_N \rho_{eq} d\Gamma = (\partial \Phi/\partial V)_{E,N}/\Omega \tag{9}$$

$$\mu \equiv \int (\partial H/\partial N)_V \rho_{eq} d\Gamma = -(\partial \Phi/\partial N)_{E,V}/\Omega \tag{10}$$

where $\Phi(E, V, N) \equiv \int \theta(E - H) d\Gamma$ is the total number of accessible states in the $\Gamma$ phase space, energetically restricted via the Heaviside step-function $\theta$. Note that $(\partial \Phi/\partial E)_{V,N} = \Omega$ since the first derivative of the Heaviside step-function is the Dirac delta function. Comparing Eqs. (9) and (10) with the relevant thermodynamic expressions $p = T(\partial S/\partial V)_{E,N}$ and $\mu = -T(\partial S/\partial N)_{E,V}$ yields the equilibrium Boltzmann's entropy (Hilbert, Hänggi, Dunkel 2014)

$$S_B = k_B \ln \Phi \tag{11}$$

which is evidently proportional to the Hartley function. As is seen, all conjugated thermodynamics parameters are easily calculated via derivatives of Boltzmann's entropy (11), because it is the characteristic function for isolated systems. The corresponding temperature $k_B T \equiv \Phi/\Omega$ obeys the exact thermodynamic relation $T = (\partial E/\partial S)_{V,N}$ as well. Since Eq. (11) reduces to the Sackur-Tetrode equation for ideal gas, the maximal thermodynamic probability $W$ is equal to $\Phi = k_B T \Omega$. Interpreting $1/\Phi$ as the uniform full probability at equilibrium, Eq. (11) shows equal probability of all system states within the Γ phase space upper bounded by the energy $E$. This should not be confusing with the uniformity of $\rho_{eq}$ along the energy surface noticed by Boltzmann.

In scientific textbooks (Huang 1991) $S_G$ is traditionally considered equivalent to $S_B$ up to a meaningless constant, which is thermodynamically unobservable. Perhaps this is reasonable at equilibrium, although $\Phi/\Omega = 1/\delta(0) = 0$ means formally zero temperature. The problem of Eq. (4) becomes even more fundamental in non-equilibrium conditions. The dynamics of the probability density in isolated systems is governed by the Liouville equation (Prigogine 1962)

$$\dot{\rho} = \{H, \rho\} = -i\hat{L}\rho \tag{12}$$

where the Poisson brackets are expressed further as the Liouville operator $i\hat{L}$, which is Hermitian. One can easily calculate now the rate of change of the Gibbs entropy (4) in isolated systems

$$\dot{S}_G/k_B = -\int \dot{\rho} \ln \rho \, d\Gamma = \int i\hat{L}\rho \ln \rho \, d\Gamma = -\int i\hat{L}\rho d\Gamma = \int \dot{\rho} d\Gamma = 0 \tag{13}$$

Therefore, the Gibbs entropy (4) does not grow in time in isolated systems, which contradicts the Second Law of thermodynamics. Quantum mechanics does not change this statement. A probable explanation of this discrepancy is that Shannon's function (4) is not applicable to the case of a microcanonical Gibbs ensemble held at constant energy, which is evident from the fact that Boltzmann's entropy grows over time (Jaynes 1965). Moreover, Eq. (13) holds for any entropy $S \equiv \int g(\rho) d\Gamma$ being simply a functional of the probability density. Therefore, $S_B$ cannot be presented as in such a form. As demonstrated by Boltzmann, the Second Law of thermodynamics requires assumption of the molecular chaos hypothesis, while the factorization of $\rho$ to a product of single particle Boltzmann's distributions $f$ is not straightforward, in general.

Finally, the fascinating objects of modern physics point out new features of entropy. For example, according to the Bekenstein-Hawking formula (Bekenstein 2004) the entropy of black holes is proportional to the area of the event horizon. Other authors (Tsekov 2022) have deduced negative temperature inside black holes, which puts in question the validity of the Third Law of thermodynamics as well. Negative temperatures are possible in systems with bounded energy spectra (Ramsey 1956) and black holes are certainly one of them. Furthermore, social entropy is identified with liberty and economic freedom in sociophysics, which points out the Second Law of thermodynamics as the driving force of the social arrow of time as well (Tsekov 2023).